%
%
\newcount\equationno      \equationno=0
\newcount\itemno  \itemno=0
\newtoks\chapterno \xdef\chapterno{}
\newdimen\tabledimen  \tabledimen=\hsize
%
%
%
\def\eqn{\eqno\eqname}
\def\eqname#1{\global \advance \equationno by 1 \relax
\xdef#1{{\noexpand{\rm}(\chapterno\number\equationno)}}#1}
%
%
%
%
\def\table#1#2{\tabledimen=\hsize \advance\tabledimen by -#1\relax
\divide\tabledimen by 2\relax\vskip 1pt
\moveright\tabledimen\vbox{\tabskip=1em plus 4em minus 0.9em
\halign to #1{#2}} }
%
%
%
\def\tablec#1#2{\tabledimen=\hsize \advance\tabledimen by -#1\relax
\divide\tabledimen by 2\relax\vskip 1pt
\moveright\tabledimen\vbox{\tabskip=1em plus 4em minus 0.9em
\halign to #1{&\hfil##\hfil\crcr #2}} }

%
\def\tablev#1#2#3{\tabledimen=\hsize \advance\tabledimen by -#1\relax
\divide\tabledimen by 2\relax\vskip 1pt
\moveright\tabledimen\vbox{\offinterlineskip\tabskip=0em
\halign to #1{\vrule##\tabskip=1em plus 4em minus 0.9em&\strut#2
&\vrule##\tabskip 0 em\crcr #3}} }
%

%
\def\boxit#1{\vbox{\hrule\hbox{\vrule\kern3pt
\vbox{\kern3pt#1\kern3pt}\kern3pt\vrule}\hrule}}
%
\def\sqr#1#2{{\vcenter{\vbox{\hrule height .#2pt
 \hbox{\vrule width .#2 pt height #1 pt \kern#1pt\vrule width .#2pt}
 \hrule height .#2pt}}}}

%
\def\beginitem{\global \advance \itemno by 1 \relax
\item{\bf \number\itemno.\hfil}}
%


%
\def\utw{\smash{\rlap{\lower5pt\hbox{$\sim$}}}}
\def\udtw{\smash{\rlap{\lower6pt\hbox{$\approx$}}}}

%

%



%



\def\bbbc{{\mathchoice {\setbox0=\hbox{$\displaystyle\rm C$}\hbox{\hbox
to0pt{\kern0.4\wd0\vrule height0.9\ht0\hss}\box0}}
{\setbox0=\hbox{$\textstyle\rm C$}\hbox{\hbox
to0pt{\kern0.4\wd0\vrule height0.9\ht0\hss}\box0}}
{\setbox0=\hbox{$\scriptstyle\rm C$}\hbox{\hbox
to0pt{\kern0.4\wd0\vrule height0.9\ht0\hss}\box0}}
{\setbox0=\hbox{$\scriptscriptstyle\rm C$}\hbox{\hbox
to0pt{\kern0.4\wd0\vrule height0.9\ht0\hss}\box0}}}}
\def\bbbe{{\mathchoice {\setbox0=\hbox{\smalletextfont e}\hbox{\raise
0.1\ht0\hbox to0pt{\kern0.4\wd0\vrule width0.3pt height0.7\ht0\hss}\box0}}
{\setbox0=\hbox{\smalletextfont e}\hbox{\raise
0.1\ht0\hbox to0pt{\kern0.4\wd0\vrule width0.3pt height0.7\ht0\hss}\box0}}
{\setbox0=\hbox{\smallescriptfont e}\hbox{\raise
0.1\ht0\hbox to0pt{\kern0.5\wd0\vrule width0.2pt height0.7\ht0\hss}\box0}}
{\setbox0=\hbox{\smallescriptscriptfont e}\hbox{\raise
0.1\ht0\hbox to0pt{\kern0.4\wd0\vrule width0.2pt height0.7\ht0\hss}\box0}}}}
\def\bbbq{{\mathchoice {\setbox0=\hbox{$\displaystyle\rm Q$}\hbox{\raise
0.15\ht0\hbox to0pt{\kern0.4\wd0\vrule height0.8\ht0\hss}\box0}}
{\setbox0=\hbox{$\textstyle\rm Q$}\hbox{\raise
0.15\ht0\hbox to0pt{\kern0.4\wd0\vrule height0.8\ht0\hss}\box0}}
{\setbox0=\hbox{$\scriptstyle\rm Q$}\hbox{\raise
0.15\ht0\hbox to0pt{\kern0.4\wd0\vrule height0.7\ht0\hss}\box0}}
{\setbox0=\hbox{$\scriptscriptstyle\rm Q$}\hbox{\raise
0.15\ht0\hbox to0pt{\kern0.4\wd0\vrule height0.7\ht0\hss}\box0}}}}
\def\bbbt{{\mathchoice {\setbox0=\hbox{$\displaystyle\rm
T$}\hbox{\hbox to0pt{\kern0.3\wd0\vrule height0.9\ht0\hss}\box0}}
{\setbox0=\hbox{$\textstyle\rm T$}\hbox{\hbox
to0pt{\kern0.3\wd0\vrule height0.9\ht0\hss}\box0}}
{\setbox0=\hbox{$\scriptstyle\rm T$}\hbox{\hbox
to0pt{\kern0.3\wd0\vrule height0.9\ht0\hss}\box0}}
{\setbox0=\hbox{$\scriptscriptstyle\rm T$}\hbox{\hbox
to0pt{\kern0.3\wd0\vrule height0.9\ht0\hss}\box0}}}}
\def\bbbs{{\mathchoice
{\setbox0=\hbox{$\displaystyle     \rm S$}\hbox{\raise0.5\ht0\hbox
to0pt{\kern0.35\wd0\vrule height0.45\ht0\hss}\hbox
to0pt{\kern0.55\wd0\vrule height0.5\ht0\hss}\box0}}
{\setbox0=\hbox{$\textstyle        \rm S$}\hbox{\raise0.5\ht0\hbox
to0pt{\kern0.35\wd0\vrule height0.45\ht0\hss}\hbox
to0pt{\kern0.55\wd0\vrule height0.5\ht0\hss}\box0}}
{\setbox0=\hbox{$\scriptstyle      \rm S$}\hbox{\raise0.5\ht0\hbox
to0pt{\kern0.35\wd0\vrule height0.45\ht0\hss}\raise0.05\ht0\hbox
to0pt{\kern0.5\wd0\vrule height0.45\ht0\hss}\box0}}
{\setbox0=\hbox{$\scriptscriptstyle\rm S$}\hbox{\raise0.5\ht0\hbox
to0pt{\kern0.4\wd0\vrule height0.45\ht0\hss}\raise0.05\ht0\hbox
to0pt{\kern0.55\wd0\vrule height0.45\ht0\hss}\box0}}}}
\def\bbbz{{\mathchoice {\hbox{$\sans\textstyle Z\kern-0.4em Z$}}
{\hbox{$\sans\textstyle Z\kern-0.4em Z$}}
{\hbox{$\sans\scriptstyle Z\kern-0.3em Z$}}
{\hbox{$\sans\scriptscriptstyle Z\kern-0.2em Z$}}}}
\def\qed{\ifmmode\sq\else{\unskip\nobreak\hfil
\penalty50\hskip1em\null\nobreak\hfil\sq
\parfillskip=0pt\finalhyphendemerits=0\endgraf}\fi}
%

\magnification=\magstep1
\hsize=5.5 true in

\centerline{INITIAL DATA AND THE END STATE OF SPHERICALLY SYMMETRIC} 
\centerline{GRAVITATIONAL COLLAPSE}

\vfill
\centerline{\bf P. S. Joshi and I. H. Dwivedi}
\centerline{\bf Tata Institute of Fundamental Research}
\centerline{\bf Homi Bhabha Road, Bombay 400 005}
\centerline{\bf India}
\vfill

\vfil\eject

\magnification=\magstep1  
\hoffset=0 true cm        
\hsize=6.0 true in        
\vsize=8.5 true in        
\baselineskip=24 true pt plus 0.1 pt minus 0.1 pt 
\overfullrule=0pt         

\centerline{\bf Abstract}

Generalizing earlier results on the initial data and the final
fate of dust collapse, we study here the relevance of the initial state of 
a spherically symmetric matter cloud towards determining its end state in the 
course of a continuing gravitational collapse. It is shown that given an 
arbitrary regular distribution of matter at the initial epoch, there always
exists an evolution from this initial data which would result either in a 
black hole or a naked singularity depending on the allowed choice of free 
functions available in the solution. It follows that given any initial 
density and pressure profiles for the cloud, there is a non-zero measure set
of configurations leading either to black holes or naked singularities, 
subject to the usual energy conditions ensuring the positivity of energy 
density. We also characterize here wide new families of black hole solutions 
resulting from spherically symmetric collapse without requiring the 
cosmic censorship assumption.

\vfil\eject

\magnification=\magstep1  
\hoffset=0 true cm        
\hsize=6.5 true in        
\vsize=9 true in        
\baselineskip=24 true pt plus 0.1 pt minus 0.1 pt 
\overfullrule=0pt         

\noindent{\bf 1. INTRODUCTION}

The role of initial data towards determining the end state of 
gravitational collapse of a spherically symmetric dust cloud has been 
analyzed in some detail recently [1]. It is understood now, for example,
that given any initial density profile for such a cloud, one can always choose
the velocity function describing the infall of the matter shells in
such a manner that either a black hole or a naked singularity will result
as the final state for such a collapse. Such results, examining the 
relevance of the regular initial data, from which the collapse 
of the cloud commences, towards its evolution in terms of a black hole
or a naked singularity should lead us to a better understanding of the 
genericity and stability aspects of naked singularities. This is most 
essential and will have important implications on any possible 
formulations of the cosmic censorship hypothesis (see e.g. [2] for a review 
of the recent developments). Even if evolutions permitting naked singularity 
were allowed by general relativity, if these were non-generic in some 
well-defined sense, that could lead to a suitable formulation for the cosmic 
censorship principle.

Our purpose here is to generalize the results such as those in [1] to
collapsing clouds with a more general form of initial data and 
matter. While dust could possibly be regarded as a good approximation 
to the state of matter in the final stages
of collapse [3], it is an idealized form  which does not take 
into account the pressures and stresses, which could play an important role
towards determining the end state of gravitational collapse. 
We consider here general type I
matter fields [4], which include most of the physically important forms 
such as dust, perfect fluids, massless scalar fields and so on. 
In fact, almost all observed forms of matter and equations of state 
would fall within this general class. Having
defined the regular initial data for the matter cloud in terms of its initial
density and pressures on an initial spacelike surface, we examine to what
possible final states the collapse of such a cloud could develop, while being
subject to the usual energy conditions requiring the positivity of energy
density. It is shown that given any regular distribution of matter
at the initial epoch, there always exists an evolution from 
this initial data which could result either into a black hole or a naked 
singularity, depending on the allowed choice of free functions available 
in the solution. We also consider here a model describing collapsing shells, 
which reveals some interesting features of gravitational collapse, and the
structure of the singularity when it is visible.

Further, our analysis here identifies wide new families of black hole 
solutions in terms of the parameters given in the initial data space or the
allowed evolutions considered. Since the cosmic censorship hypothesis is 
widely useful for black hole physics, it is of crucial importance to isolate 
and identify the causes in the gravitational collapse phenomena in general
relativity which may be responsible for the formation of naked singularities
or black holes. The role of the initial data in this context 
can not be overemphasized. In fact, it is already known in several 
recent naked singularity examples that the initial data at the onset of the 
collapse does categorize the naked singular or black hole spacetimes, 
resulting from collapse. A particularly transparent example, apart from the 
dust collapse, is that of the Vaidya-Papapetrou radiation collapse models, 
where the rate of implosion of the mass ($\lambda\equiv dm(u)/du$ before the
formation of singularity determines whether the collapse would end in a naked 
singularity or a black hole (see e.g. [5] for a discussion).  
For such collapse models, there exist 
non-zero measure sets in the space of initial data, which evolve in either 
a naked singularity or a black hole.

In other words, given the initial matter profiles in terms of the 
densities and pressures distributions, our results here indicate a specific
procedure to fine tune the evolutions so as to necessarily produce a 
black hole as the end product of collapse. Such a conclusion no longer
requires the assumption of cosmic censorship, which is turning out to be
a difficult proposition to establish rigorously. 
From such a perspective, ours is a 
more specific characterization, in terms of the space of initial data and the
possible evolutions, in order to generate black holes.

\bigskip

\noindent {\bf 2. EINSTEIN EQUATIONS AND REGULARITY CONDITIONS}

We are interested here in the problem of how a given regular initial data 
set evolves dynamically in a spherically symmetric spacetime in the context
of the occurrence and nature of singularities. Our main purpose is
to analyze the Einstein field equations, with a given initial data set such as
the state of matter and the velocities of the
spherical shells at the onset of collapse for a compact object, 
in order to determine the possibilities of this configuration evolving 
into either a black hole or a naked singularity.
We therefore consider the gravitational collapse of 
a matter cloud that evolves from
a regular initial data defined on an initial spacelike surface.
The energy-momentum tensor has a compact support on this initial
surface where all the physical quantities such as densities and pressures are 
regular and finite. For such a cloud of sufficiently high total mass, 
there will not be any stable equilibrium configurations available, 
and a continual
gravitational collapse must ensue resulting finally into a 
space-time singularity of infinite densities and curvatures.

For the matter distribution of the spacetime within the cloud
undergoing gravitational collapse, we shall not restrict ourselves to 
any specific form of matter but consider general type I matter fields.
In fact, all the known and observed
physical matter fields are of this type, except the directed radiations 
which are of type II. It thus follows that as far as the form of the matter is
concerned, we have included quite general collapse scenarios
which include all the physically reasonable
matter fields, and possible equations of state. The type I and type II 
matter fields 
are characterized by the existence of two real orthogonal eignvectors, 
or one double-null real eignvector respectively, in a timelike invariant 
2-plane in the spacetime. The remaining fields of type III and type IV are 
considered physically unreasonable, as they 
necessarily violate the energy conditions ensuring the positivity of the 
mass-energy density. Furthermore, such fields have not been observed 
in nature so far, all the observed matter fields being of type I. 
Thus, we would not attribute any physical significance or
interpretation to the same presently. A type II 
matter distribution corresponds to zero-rest mass fields representing 
a directed radiation. The radiation collapse, corresponding to such a field 
can be described by a Vaidya spacetime and has been 
studied in detail to show that both the black holes and
naked singularities could result in such a collapse.

Thus we concentrate here on type I matter fields,
for spherically symmetric spacetimes. Such a matter field, in a general
coordinate system, can be expressed as
$$T^{ab}=\lambda_1 E^a_1E^b_1 +\lambda_2 E^a_2E^b_2
+\lambda_3 E^a_3E^b_3 +\lambda_4 E^a_4E^b_4\eqn\qq$$
where (${E_1,E_2,E_3,E_4}$) is an orthonormal basis with ${E_4}$ and $(E_1,
E_2,E_3)$
being timelike and spacelike eignvectors respectively, with
$\lambda_{i}$s ($ i=1,2,3,4$) being the eigenvalues.
For such a spherically symmetric matter distribution
we can choose coordinates $(x^i=t,r,\theta,\phi)$ adopted to the above
orthonormal frame, and the metric is written as,
$$ds^2=-e^{2\nu}dt^2+e^{2\psi}dr^2+R^2d\Omega^2\eqn\qq$$
where $d\Omega^2= d\theta^2+ \sin^2\theta d\phi^2$ is the line element
on a two-sphere. 
Here $\nu,\psi$ and $R$ are functions of $t$ and $r$, and the stress-energy
tensor $T^a_b$  given by equation (1) 
has only diagonal components in this coordinate system (i.e. we are
using a comoving coordinate system), given by 
$$T^t_t=-\rho,\quad T^r_r=p_r,\quad T^{\theta}_{\theta}=p_{\theta}=
T^{\phi}_{\phi},\quad T^t_r=T^r_t=0\eqn\qq$$
The quantities 
$\rho, p_r$,  and $p_{\theta}$ are the eigenvalues of $T^a_b$ and are
interpreted as the density, radial pressure, and tangential stresses 
respectively.
We take the matter fields to satisfy the weak energy condition,
i.e. the energy density as measured by any local observer must be
non-negative, and so for any timelike vector $V^a$ we must have 
$$ T_{ab}V^aV^b\ge 0\eqn\qq$$
which amounts to
$$\rho\ge0,\quad \rho+p_r\ge0,\quad \rho+p_{\theta}\ge 0\eqn\qq$$

From the point of view of the dynamical evolution of the initial
data at an epoch of time from which the collapse commences, one  
has a total of six arbitrary functions of $r$, namely
$$\nu(t_i,r)=\nu_o(r),\quad \psi(t_i,r)=\psi_o(r),\quad R(t_i,r)=R_o(r),$$
$$\rho(t_i,r)=\rho_o(r),\quad p_r(t_i,r)=p_{r_o}(r),\quad p_{\theta}(t_i,r)
=p_{\theta_o}(r)\eqn\qq$$
These functions constituting the initial data are to be specified at 
some initial spacelike surface at an initial time $t=t_i$.
The dynamical evolution of such a set of initial data is determined by the
Einstein equations, and for the metric (2) these are given by,
$$ T^t_t=-\rho=-{F'\over k_oR^2R'},\quad
T^r_r=p_r=-{\dot F\over k_oR^2 \dot R}\eqn\qq$$
$$\nu'(\rho +p_r)=
2(p_{\theta}-p_r){R'\over R}-p_r'\eqn\qq$$
$$-2\dot R'+R'{\dot G\over G}+\dot R {H'\over H}=0\eqn\qq$$
$$ G-H=1-{F\over R}\eqn\qq$$
where $(\dot{ })$ and $( ' )$ represent partial derivatives with respect to
$t$ and $r$ respectively, $F=F(t,r)$ is an arbitrary function of $t$ and $r$, 
and we have put
$$G=G(t,r)=e^{-2\psi}(R')^2,\quad H=H(t,r)=e^{-2\nu}\dot R^2\eqn\qq$$

The initial data represented by the functions $\nu_o,\psi_o,\rho_o,
p_{r_o},p_{\theta_o}$ and $R_o$ are not all independent. 
From equation (8) it follows that there is a relation between them as given by
$$\nu_o'(\rho_o +p_{r_o})=
2(p_{\theta_o}-p_{r_o}){R'_o\over R}-p_{r_o}'\eqn\qq$$
$$\Rightarrow \nu_o(r)=\int {\left({(2p_{\theta_o}-2p_{r_o})R_o'\over R_o(
\rho_o +p_{r_o})}-{p_{r_o}'\over \rho_o +p_{r_o}}\right)dr}\eqn\qq$$
Furthermore, there is a coordinate freedom left in the choice of 
the scaling of the coordinate $r$, which can be used to reduce the 
number of independent initial data to four. It follows that 
there are only four independent arbitrary functions of $r$  
constituting the initial data.
Evolution of this data is governed by the field equations,
and we have in all five equations with seven unknowns,
namely $\rho,p_r,p_{\theta},\nu,\psi,R$ and $F$, giving us freedom of choice
of two functions. Selection of these two free functions, subject to the given
initial data and the weak energy condition above, 
determines the matter distribution and the metric of the 
spacetime  and thus leads to a particular evolution for the initial data.

In spherically symmetric spacetimes, the function $F(t,r)$ is treated 
as the mass function for the cloud with $F\ge 0$. In order to preserve the
regularity of the initial data at $t=t_i$, we must have $F(t_i,0)=0$, that
is, the mass function vanishes at the center of the cloud.

In this paper, we are interested only in the spacetimes which 
at some initial epoch $t=t_i$ are free from any singularities in the initial
data, i.e. the evolution of the collapse must develop from a 
regular initial data.   
Our interest is in the evolutions of such data sets, 
which develop from this initial singularity free state, into a 
possible singularity of 
the spacetime. We then study the nature and structure of this singularity 
from the perspective of the cosmic censorship. 
The initial data is considered to be singularity free if the curvatures
and densities are all finite, that is the Kretchmann scalar is bounded
on the initial surface.
The Kretchmann scalar $K=R^{abcd}R_{abcd}$
for a spherically symmetric
spacetime, as given by the metric in (2), can be put in the following form,
$$K=C^2-{1\over 3}T^2+2{\bf T}^2\eqn\qq$$
where
$$C^2=C^{abcd}C_{abcd}={4\over 3}({F\over R^3}+p_r-p_{\theta}-\rho)^2\eqn\qq$$
$$T^2=(T^a_a)^2=(p_r+2p_{\theta}-\rho)^2\eqn\qq$$
$${\bf T}^2=T^{ab}T_{ab}=\rho ^2+p_r^2+2p_{\theta}^2\eqn\qq$$
Thus, a singularity will appear on the initial surface 
if either the density $\rho$, or 
one of the pressures become unbounded at any point on the initial 
surface, or $(F/R^3)\rightarrow \infty$ at any point. 
We require that at the initial surface $t=t_i$, the density and pressures 
are finite and bounded. Further more, we have 
$$F(t_i,r)=\int{\rho_o(R_o)R^2_odR_o}\eqn\qq$$
and hence the spacetime is singularity free initially in the sense that
the Kretchmann scalar, density and pressures are finite. But as the
collapse evolves, a singularity could develop at a later time whenever 
either of the density or one of the pressures become unbounded. 
We shall consider below such specific evolutions of the initial data
which model a gravitationally collapsing matter cloud.
\bigskip

\noindent{\bf 3. COLLAPSING MATTER CLOUDS}

The initial data for a collapsing matter cloud basically means the
initial densities and pressures describing the initial state of matter 
at the onset of collapse, namely, $\rho_o,p_{r_o}, p_{\theta_o}$,
and the function $\psi_o$, which is related to the initial velocity of 
the collapsing shells. From the point of view of the cosmic censorship
hypothesis, all regular, physically reasonable initial data sets should evolve
into a black hole. It is known, of course, that there are initial
data sets with reasonable forms of matter such as the Lemaitre-Tolman-Bondi
dust collapse, radiation collapse, and self-similar adiabatic perfect 
fluid scenarios, which result into either a black hole or a naked singularity,
depending on the nature of initial data and initial distributions of
densities and pressures. Thus, the above broad concept of cosmic censorship
has to be fine tuned and made more precise in order to arrive at a genuine
formulation and proof for censorship (see e.g. [2] or [6]).

It follows that either for such a purpose of cosmic censorship, or to 
understand the naked singularities of gravitational collapse better, 
we need to examine the evolutions of an arbitrary, but regular and physically 
reasonable set of initial data undergoing the gravitational collapse. 
This requires defining more precisely as to what constitutes
a regular, physically reasonable, initial data. As far as the form of the 
matter is concerned, since all the observed forms of matter have been
of type I as discussed earlier, we have considered here general type I 
matter fields. Next, it is widely accepted that all physically reasonable
matter forms must satisfy the energy conditions ensuring the positivity
of mass-energy densities. Therefore, one of the
conditions we must impose naturally is that at the onset of the collapse, 
all the initial
data sets specifying the density and pressures profiles of the cloud
must satisfy an energy condition, and the same must hold during
the later evolution of collapse. 
The regularity of the initial data also means that at 
the onset of collapse at some $t=t_i$, the spacetime must be singularity
free, as discussed earlier. Further more,
from the point of view of general relativity the data should be at least
$C^2$ differentiable. Therefore we require that the functions 
$\rho_o, p_{r_o}, p_{\theta_o}$ be atleast $C^2$ function of $r$. 
The same applies to the metric functions $\nu_o$ and $\psi_o$ 
which must be $C^2$ functions of $r$ throughout the spacelike section  
$t=t_i$ for all $r\ge 0$.

As stated above, for the initial data to be physically reasonable
the weak energy condition is to be satisfied. This imposes the following 
restrictions on the initial data,
$$\rho_o(r) \ge 0,\quad \rho_o+p_{r_o}\ge 0,\quad \rho_o(r)+
p_{\theta_o}(r)\ge 0\eqn\qq$$
Further, for the evolution of the matter in the spacetime to be 
physically realistic, energy conditions (5) have to be satisfied.
Therefore we assume that through out the cloud, at the initial and at 
all later epochs the weak energy condition is satisfied during
the entire evolution of the collapsing cloud till the formation of the 
spacetime singularity at $R=0$.

Since the function 
$\nu_o$ is related to the initial matter data by equation (13), 
and is a $C^2$ function of $r$, this puts certain restrictions on the 
arbitrariness of the choice of the functions $\rho_o,p_{r_o},p_{\theta_o}$.
For example, we have
$$[p_{\theta_o}-p_{r_o}]_{r=0}=0 \eqn\qq$$
For the sake of physical reasonableness, we require the center $r=0$ to be 
the regular center for the cloud, which means $R(t,0)=0$.  
Also, one would like to have the initial density $\rho_o(0)>0$ 
at the center $r=0$. This implies that we have
$$\rho_o(0) +p_{r_o}(0)>0,\quad  \nu_o(r)=r^2h(r)\eqn\qq$$
where $h(r)$ is at least a $C^1$ function of $r$ for $r=0$, and at least
a $C^2$ function for $r>0$.
This means that the pressure gradients 
vanish at the center $r=0$, basically meaning that the forces vanish 
at the center. In this section we consider the above scenario for the 
sake of physical reasonableness, however, it is possible to give a more 
general formalism independent of requirements such as above, as we shall 
discuss in section 6.

In fact, it would be reasonable to require that pressures be positive 
at the onset of the collapse, since for astrophysical bodies physically 
we would prefer the pressures rather than tensions. 
Further more, to make the scenario physically more appealing, we may 
require the density to be decreasing as we move away from the center $r=0$.
In that case, for various reasonable equations of state such as 
$p=k\rho, 0<k<1,$ (a perfect fluid), or $p=k\rho^\gamma$, the pressure also
will decrease away from the center together with the decreasing density.
This will be typically the case in the massive bodies such as stars and such
other astrophysical systems. Then as such the energy conditions  
could impose restrictions on the maximum size of the matter cloud with 
such an initial density and pressure distribution.

We use the coordinate freedom available in rescaling the radial 
coordinate $r$ such that
$$R(t_i,r)=R_o(r)=r\eqn\qq$$
The physical area radius $R$ then monotonically increases with the 
coordinate $r$, and there are no shell-crossings on the initial surface,
with $R'=1$. Since we are considering gravitational collapse,
we also have $\dot R<0$.

Consider now the gravitational collapse of a matter cloud with a 
general initial data as prescribed above. We supply the two free functions
$F$ and $\nu$ as below,
$$\nu=c(t)+\nu_o(R),\quad F=f(r)+F_o(R)\eqn\qq$$
From equations (7) to (13), the evolution of the collapse 
is described by the equations,
$$\nu_o(R)=R^2g(R)=\int_0^R {\left({2p_{\theta_o}-2p_{r_o}\over r(
\rho_o +p_{r_o})}-{p_{r_o}'\over \rho_o +p_{r_o}}\right)dr}\eqn\qq$$
$$G=b(r)e^{2\nu_o}\eqn\qq$$
$$\sqrt{R}\dot R=-a(t)e^{\nu_o}\sqrt{b(r) Re^{\nu_o}-R+f+F_o}\eqn\qq$$
$$\rho={f'\over R^2R'}+{F_o,_R\over R^2}\eqn\qq$$
$$p_r=-{F_o,_R\over R^2}\eqn\qq$$
$$2p_{\theta}=R\nu,_R(\rho +p_r)+2p_r+Rp_r,_R\eqn\qq$$
$$F_o(R)=-\int_0^R{r^2p_{r_o}dr}\equiv -R^3{\it p}(R),\quad 
f(r)=\int_0^r{r^2(\rho_o+p_{r_o})dr}\equiv
r^3\epsilon(r)+r^3p(r)\eqn\qq$$

Here $F(t_i,r)=2r^3\epsilon(r)$. The quantities $\epsilon(r)$ and 
$p(r)$ are to be treated 
as the average mass and pressure densities of the cloud, and are 
decreasing functions of $r$. Since $p_{r_o}$ is a positive function on
the initial surface, it follows that $F_o<0$ throughout the spacetime, 
and as such the radial pressure is non-negative throughout the spacetime.
The arbitrary function $b(r)$ characterizes the velocity of the spherical 
shells at the initial time $t=t_i$. We are dealing with the collapse 
situation with $\dot R <0$, therefore the arbitrary function $a(t)>0$.

It should be noted that for the case $p_r=p_\theta=0$, the above set 
of equations, and the collapse models reduces to the Tolman-Bondi-Lemaitre 
case of general inhomogeneous dust collapse. In that case, the results in [1]
apply as far as the role of the initial data towards determining the final 
fate of collapse is concerned in terms of either a black hole or a naked
singularity. Thus, the models here may also be  
viewed as directly generalizing
the Tolman-Bondi-Lemaitre results to include both the radial and
tangential pressures in order to investigate again the role of initial
data towards the final fate of collapse. 
Certain examples have been examined which include non-zero pressures,
and it is seen that again both black holes and naked singularities arise
as final fate of collapse [7].

Since $r=0$ is the regular center of the cloud meaning $R(t,0)=0$,
it follows from equation (26) that 
$$\sqrt{v}\dot v=-a(t)e^{\nu_o}\sqrt{v^3(h(R)b(r)-{\it p}(R))+ b_o(r)v+
\epsilon(r)+{\it p}(r)}\eqn\qq$$
where the arbitrary function $b(r)=1+r^2b_o(r)$, such that $b_o(r)$ is 
at least a $C^1$ function of $r$ for $r=0$, and a $C^2$ function for $r>0$, 
and we have put
$$R=rv(t,r),\quad v(t_i,r)=1\eqn\qq$$
$$h(R)=h(rv)={e^{r^2v^2g(rv)}-1\over r^2v^2}\eqn\qq$$
The functions $b_o(r), h(rv), v(t,r), u(rv)$, and $f_o(r)$ are all at least
$C^1$ functions of their arguments. Note that at $t=t_i$ we have 
$v=1$ and since $\dot v <0$, we have $v<1$ throughout the spacetime.

We note that the quantity $R(t,r)\ge 0$ here is the area radius 
in the sense that $4\pi R^2(t,r)$ gives the proper area 
of the mass shells at any given value of the
comoving coordinate $r$ for a given epoch of time. The area of such a 
shell at $r=$const. goes to zero when $R(t,r)=0$. In this sense, the curve 
$t=t_s(r)$ such that $R(t_s,r)=0$
describes the singularity in the spacetime where the mass shells are
collapsing to a vanishing volume, where the density and pressures
diverge. This shell-focusing singularity occurs along the curve
$t=t_s(r)$ such that $v(t_s,r)=0$, the Kretchmann scalar diverges
at such points. Using the remaining degree of freedom left
in the scaling of the time coordinate $t$ we could put $a(t)=1$. 
Equation (31) can then be integrated with the initial condition $v(t_i,r)=1$ 
to obtain the function $v(t,r)$.
Note that the coordinate $r$ is treated as a constant in the equation. 
We in fact get
$$\int_v^1{\sqrt{v}dv\over \sqrt{b_o(r)ve^{3\nu_o}+e^{2\nu_o}(v^3(h(rv)-p(rv))
+\epsilon(r)+{\it p}(r))
}}=t
\eqn\qq$$
where we have chosen for the sake of simplicity $t_i=0$. The time $t=t_s(r)$ 
corresponds to the occurrence of singularity is then given by,
$$t=t_s(r)=\int_0^1{\sqrt{v}dv\over 
\sqrt{b_o(r)ve^{3\nu_o}+e^{2\nu_o}(v^3(h(rv)-p(rv))
+\epsilon(r)+{\it p}(r))
}}
\eqn\qq$$

For such a collapsing cloud, a singularity can also occur at 
points where $R'=0$, which are termed the shell-crossing singularities.
Though these are singularities of a weaker nature in general through which
the spacetime can possibly be extended using a suitable extension
procedure [8], the comoving coordinate system we have used
here may break down and the metric might possibly become degenerate 
at the points where $R'=0$. However, our purpose here is to study the 
shell-focusing singularity at $R=0$, which is essentially different and
could be much stronger gravitationally as compared to the shell-crossings 
which are possibly delta-function like singularities, caused by different 
shells crossing each other where the density momentarily blows up. 
Hence, we choose the evolution of the initial data in such a 
manner that any shell-crossings are avoided in the collapse, except
possibly at the singularity. Here it may be useful to mention that
a similar situation regarding the occurrence of shell-crossings arises 
in Tolman-Bondi-Lemaitre dust collapse models also. However, as
has been pointed out in earlier works, for a given initial density
profile one can always choose appropriate initial velocity of the
dust shells such that during the evolution no shell-crossings are
encountered or visa-versa.

In a similar fashion, this can be achieved in the present general case 
as well by a suitable choice of the functions involved,
as specified below. At a given epoch of time, the functions $\epsilon (r)$, 
$p(r)$, $\nu_o(R)$ and $h(R)$ are at least $C^1$ functions and further more 
$\epsilon(r)$ and $p(r)$ are decreasing functions of $r$. The function 
$b_o(r)$ is an arbitrary function representing the initial velocities 
of the collapsing spherical shells. From equation (35) it is clear that 
the singularity time $t_s(r)$ is an explicit
function of the velocity function $b_o(r)$ which is a free function,
and one can choose it in such a way that $t_s(r)$ is
an increasing function of $r$, i.e. $dt_s/dr >0$. The exact nature of such
velocity functions $b_o(r)$ for which $t_s(r)$ in equation (35) is an
increasing function of $r$, depends upon the exact behavior of initial density
and pressures within the cloud. For example, for a matter
cloud initially satisfying an equation of state of the type $p=a\rho^{\gamma}$
one of the many possibilities is the function $b_o(r)>0$ such
that $b_o'(r)<0$ and is less than a certain minimum for $r_b\ge r\ge 0$. 
For all such functions 
$b_o(r)$, therefore, the singularity curve $t=t_s(r)$ is an increasing curve 
for all allowed values of coordinate $r$, and 
the successive spherical shells within the cloud
collapse to singularity successively, and shell crossings do not
occur. Thus $R'=v+rv'>0$ (note that $R'=1$ initially) throughout the spacetime.
Furthermore, if $[r^2b_o]'\ge 0$ than  $\sqrt{v}R'\le 1$ within the cloud for
$1\ge v\ge 0$.

The shell-focusing singularity $R=0$ occurs first at $r=0$ 
and the time of occurrence of such a singularity from equation (35)
is given by,
$$t_{s_o}=t_s(0)=\int_0^1{\sqrt{v}dv\over 
\sqrt{v^3(h_o-p_o)+b_{oo}v+\epsilon_o+p_o}}
\eqn\qq$$
where $h_o=h_o(0),p_o=p(0), \epsilon_o=\epsilon(0), b_{oo}=b_o(0)$ 
are constants related to the central density and pressures,
In fact, near the center $r=0$ we have
$$t_s(r)=t_{s_o}+rX(0)+ O(r^2)...\eqn\qq$$
where the function $X=X(v)$ is given by
$$X(v)=\sqrt{{\epsilon_o+p_o+b_{oo}v+v^3(h_o-p_o)\over 
\epsilon_o+p_o}}
\int_v^1{{\sqrt{v}(\epsilon_1+p_1+b_1v-v^4h_1)dv\over 
(v^3(h_o-p_o)+b_{oo}v+\epsilon_o+p_o)^{3/2}}}\eqn\qq$$
where $\epsilon_1=-\epsilon'(0), p_1=-p'(0), b_1=-b_o'(0), h_1=h,_R(0)$.
For the case that we are considering where the pressures have been taken to 
be positive, the central singularity at $r=0$ could be naked, all subsequent 
singularities with $r>0$ are covered as the quantity $F/R\rightarrow \infty$
and the trapped surfaces and the apparent horizon develop prior to the
formation of the singularity. (When the pressures are allowed to be 
negative, still subject to the validity of the weak energy condition, the 
other parts of the singularity can be visible in principle, see e.g. 
Cooperstock et al in [7]). 
It thus remains only to examine the nature of the central singularity.

We now show that the weak energy condition is satisfied by the class of
collapse models under consideration. For the models to be physically 
reasonable, it is necessary that the positivity of energy density is preserved
at the initial epoch, and also throughout  the entire evolution of the 
collapse, till the formation of the spacetime singularity at $R=0$. 
At the initial epoch, for the regularity and physical reasonableness of 
the initial data, we require the initial matter density $\rho_o(r)$ to 
be positive, and also the initial radial and tangential pressures $p_{r_o}$ 
and $p_{\theta_o}$ to be positive, or at least non-negative. Thus we have,
$$\rho_o(r)\ge0,\quad p_{r_o}\ge0, \quad p_{\theta_o}\ge0$$

The initial density $\rho_o(0)=\rho_c$ is maximum at 
the regular center $r=0$, and decreasing away from the center, and the central 
pressures $p_r,p_\theta$ are also positive. It is therefore intuitively 
clarifying to understand the behavior of these physically relevant quantities 
in the neighborhood of the regular center $r=0$, and near the singularity, 
during the evolution of the collapsing cloud. 
Firstly, it is clear from equation (28) that the radial pressure $p_r$ 
will be positive throughout out the subsequent evolution of the cloud. 
It then follows from equation (27) and (28) that $\rho +p_r\ge 0$ throughout 
the spacetime, since it is that way to begin with. Next, note that in
the near regions of the center at $r=0$, and that of the singularity
at $v=0$, we have $\rho\ge 0, \rho +p_r\ge 0, \rho+p_{\theta}\ge 0$, 
if the same is satisfied at the initial surface, regardless of any 
restrictions on the form of initial densities and pressures. 
Actually, the initial density $\rho_o(0)=\rho_c$ is maximum at 
the regular center $r=0$, and decreasing away from the center, and the central 
pressures $p_r,p_\theta$ are also positive there. 
In the near regions of the center $r=0$, 
$\sqrt{v}R'=v^{3/2}+rX(v)+O(r^2)$, $X(v)<1$,  and we have from equation
(27) to (29)
$$v^2R'\rho= \rho_c-\rho_1r+p_{r_c}(1-v^2-v^{3/2}X(v)r)+O(r^2)\eqn\qq$$
$$v^2R'(\rho+p_{\theta})=
\rho_c+p_{\theta_c}-\rho_1r{\rho_c+p_{\theta_c}\over \rho_c+p_{r_c}}+O(r^2)
\eqn\qq$$ 
where $\rho_c$, $p_{r_c}$ and $p_{\theta_c}$ represent the central density
and pressures at the initial epoch $t=t_i$, and $\rho_1$ is the first
derivative of $\rho_o(r)$ at the center. 
As $v\rightarrow 0$ we have
$\rho \rightarrow +\infty$,  $\rho +p_r \rightarrow +\infty$, and 
$\rho +p_{\theta}\rightarrow +\infty$. 
Thus $\rho\ge 0, \rho +p_r\ge 0$ and $\rho +p_\theta\ge 0$, and the energy
conditions are clearly satisfied in a certain neighborhood of the 
regular center $r=0$ through out the evolution as seen from the above.
It is seen that at the shell focusing singularity $v=0$ the density becomes 
infinite where the curvature singularity occurs. In the case when the 
densities and pressures are initially decreasing functions of $r$, then 
throughout the cloud they have maximum values at the center, and are 
decreasing within the cloud.

In fact, a sufficient condition that the energy conditions be satisfied 
for a matter cloud of a finite size $r_c\ge r\ge 0$ at all times is that 
$(f(r)-f(R))'\ge 0$ within the coordinate 
range $r_c\ge r\ge 0$. Note that the inequality is satisfied both initially
for $v=1$ (i.e. at $t=0$), in near central region throughout the evolution,
and in the neighborhood of the entire singularity curve $v=0$.
In fact, as $v\to 0$ we have $\rho\to +\infty, \rho +p_r\to +\infty,
\rho +p_{\theta}\to +\infty$. Further, it follows from equation (30) that 
for any given value of the radial coordinate $r$,
$$f(r)= \int_0^r {r^2 (\rho_o+p_{r}) dr}\ge f(R(r,t))=\int_0^R {r^2 
(\rho_o+p_{r}) dr}$$
Hence the mass function $F\ge 0$ for all $r$ at all times throughout the
cloud. Since $f'=r^2(\rho_o+p_{r0})>0$ and is an increasing function of 
$r$ for $r_c\ge r\ge 0$ where $f''(r_c)=0$, it follows that $f'(r)\ge 
f_{,R}(R)R'$, and energy conditions are satisfied within $r_c\ge r\ge 0$.
Thus, for a given initial matter distribution as above, there is a finite
cloud where the energy conditions are satisfied through out the
evolution of the initial data.
Next, for a cloud of arbitrary size and for situations with completely 
arbitrary initial density and pressures which are not
necessarily decreasing (or even including negative pressures, or pressures and
density increasing with increasing $r$ etc.), note
that the negative contribution in the expressions for $\rho$ and $\rho +
p_{\theta}$ is finite throughout the evolution while the positive terms are
maximum at the center and become unbounded at the singularity.  
The behavior of the term $R'f_{,R}=r^2v^2R'(\rho_(R)+p_r(R))$ (specificely 
$R'\propto b^{-1}_o(r)$ for $v>0$) depends explicitly on the 
velocity function $b_o(r)$ which is a free function, while $f'(r)=r^2
(\rho_o(r)+p_r(r))$ is 
independent of the same. Therefore a suitable choice of $b_o(r)$ such that 
$${\rho_r(r)+p_r(r)\over \rho_o(R)+p_r(R)}\ge v^2(v+rv')\eqn\qq$$
within the range $r_b\ge r\ge 0$ would result in the energy conditions 
being satisfied for such an arbitrary size cloud for a given initial data.

Clearly, depending upon the conditions and values on the boundary of the cloud
for the density and pressures, the exterior matching for the general
class of models given here could be Schwarzschild,
or one of the other suitable spacetimes, such as for example a Vaidya model
spacetime. We shall not
enter into these details of the matching presently, as our main concern here
is the local nakedness or otherwise of the central singularity at $R=0$.

\bigskip

\noindent{\bf 4. APPARENT HORIZON AND THE NATURE OF SINGULARITY}

The apparent horizon within the collapsing cloud is given by $R/F=1$,
which gives the boundary of the trapped surface region in the spacetime. 
It is the behavior
of the apparent horizon curve (which meets the central singularity at
$R=r=0$) near the center which essentially determines
the visibility, or otherwise, of the central singularity. For example,
it is known within the context of the Tolman-Bondi-Lemaitre models that
the apparent horizon can be either past pointing timelike or null, or it can
be spacelike, as can be seen by examining the nature of the induced metric 
on this surface. This is unlike the event horizon curve which is always 
future pointing null. If the neighborhood of the center gets trapped earlier 
than the singularity, then it is covered, and if that is not the case the
singularity can be naked [9], with families of nonspacelike trajectories
escaping from it.

In order to consider the possibility of existence of such families,
and to examine the nature of the central singularity occurring at 
$R=0, r=0$ in the general class of models considered in the previous section,
let us consider the equation of the outgoing radial null geodesics
which is given by,
$${dt\over dr}=e^{\psi-\nu}\eqn\qq$$
The singularity appears at the point $v(t_s,r)=0$ which corresponds to 
$R(t_s,r)=0$, therefore if there are future directed outgoing radial null 
geodesics, terminating in the past at the singularity, then along these 
trajectories we have $R\to 0$ as $t\to t_s$.

Writing the equation for these radial null geodesics in terms of the 
variables $(u=r^{5/3},R)$ we get, 
$${dR\over du}= {3\over 5}({R\over u}+\sqrt{v}v'{1\over {R\over u}}
){1-{F\over R}\over \sqrt{G}(\sqrt{G}+\sqrt{H})}\eqn\qq$$
If the null geodesics terminate in the past at the singularity with a 
definite tangent, then at the 
singularity the tangent to the geodesics $dR/du>0$ in the $(u,R)$ plane,
and must have a finite value.
In the case of collapsing ball of matter we are considering, all 
singularities at $r>0$ are
covered since $F/R\rightarrow \infty$, and therefore $dR/du \rightarrow 
-\infty$, and only the singularity at center $r=0$ could be naked. 
As mentioned earlier, for the case when $R'>0$ near the central singularity, 
we have
$$x_o=\lim_{t\to t_o,r\to 0}{R\over u}=\lim_{t\to t_o,r\to 0}{dR\over du}
\Rightarrow x_o^{3/2}={3\over 2}X(0)\eqn\qq$$
where $X(0)$ is given by equation (37). Since $X(0)>0$ the singularity is 
at least locally naked. The behavior
of outgoing radial null geodesics in the neighborhood of the singularity
are described by $R=x_ou$ in $(R,u)$ plane and in $(t,r)$ plane it
is given by
$$t-t_s(0)=x_or^{5/3}\eqn\qq$$

The global visibility of such a singularity will depend on the overall
behavior of the various functions concerned within the matter cloud and we
shall not go into those details presently. It has been seen, however, from the
study of various examples so far, that once the singularity is locally naked,
one can always make it globally visible by a suitable allowed choice of 
functions. Note that in cases where the choice of $b_o(r)$ is such that 
$X(0)<0$ the singularity would be covered.
When $X(0)=0$, one has to consider the next higher order expansion term
which is nonvanishing in equation (37), and that will then determine the
nature of the singularity by means of essentially a similar analysis.

One can also write the equation for these radial null geodesics
in terms of the variables $(t,R)$ to see how the area radius $R$ grows along
these outgoing null geodesics with increasing values of time.
As mentioned earlier, for the case where
$R'>0$ near the central singularity, we get
$$X_o=\lim_{t\to t_o,r\to 0}{R\over t-t_s(0)}=
\lim_{t\to t_o,r\to 0}{dR\over dt}= 
=\lim_{t\to t_o,r\to 0}[e^{\nu}{1-{F\over R}\over \sqrt{G}+\sqrt{H}}]
=1\eqn\qq$$
Again this shows that the singularity is naked at least locally. 
In fact, as pointed out above, it follows that for
$b_o'(0)\ne 0$ the area coordinate behaves as $R=const. r^{5/3}$
near the singularity. 
\bigskip

\noindent{\bf 5. COLLAPSING SHELLS}

We next consider here collapsing thick shells of matter, subject to the
weak energy condition, but otherwise with a general form of matter, as
this reveals some interesting features regarding the structure of the 
singularity. We rescale 
the coordinate $r$, contrary to the earlier case of collapsing matter clouds,
such that at some initial $t=t_i$ the density distribution of shells is 
expressed as 
$$\rho_o(r)=\rho_c(1-\rho_1r^{\alpha})\eqn\qq$$
where $\rho_c>0, \rho_1>0$ and $\alpha \ge 1$ are constants.
It is seen that in the scaling given as above, the area radius $R$ again
increases with the increasing coordinate $r$.

We choose functions $F$ and $G$ such that
$$F=f(t)R,\quad G=a(r)e^{\Phi(R)}\eqn\qq$$
where $f(t)$ and $a(r)$ are  arbitrary $C^2$ functions of $t$ and $r$ only.
From field equations (7) to (11) we get the matter distribution 
in spacetime being characterized by $\rho,p_r$ and $p_{\theta}$
$$\rho={f(t)\over R^2}\eqn\qq$$
$$p_r=-{\dot f\over R\dot R}-{f(t)\over R^2}\eqn\qq$$
$$p_{\theta}=-{\dot f\over 2R\dot R}(1-{R\over 2}\Theta,_{R})\eqn\qq$$
$$e^{\Theta}=ae^{\Phi}+1-f\eqn\qq$$
$$\dot R=-b(t)e^{{\Phi\over 2}}\sqrt{ae^{\Phi}+1-f}\eqn\qq$$
where $H=e^{\Theta}$ and $\Theta(R,t)$ is treated as function of $R$
and $t$ and $(,_R)$ represents partial derivative with respect to $R$ with
$t$ kept constant. Also, $b(t)>0$ is an arbitrary function of $t$.
If the spacetime 
characterized by the solutions of the field equations given by these equations
does indeed evolve from the initial matter data $\rho_o(r),p_{r_o}(r)$ and
$p_{\theta_o}(r)$, one must have at the initial epoch $t=t_i$
$$\rho(t_i,r)=\rho_o(r)={f_o\over R_o^2},\quad f_o=f(t_i),\eqn\qq$$
$$p_r(t_i,r)=p_{ro}={ f_1\over R_ov_o}-{f_o\over R_o^2},\quad f_1=[\dot f]_
{t=t_i},v_o(r)=-v(t_i,r),\eqn\qq$$
$$p_{\theta}(t_i,r)={ f_1\over 2R_ov_o}(1-R_o\mu,_{R_o}),\quad
2\mu(R_o)=\Theta(R_o,t_i)-\log c_o\eqn\qq$$
where $c_o$ is a constant. On further simplification we  get
$$R_o^2(r)={f_o\over \rho_c-\rho_1r^{\alpha}}\eqn\qq$$
$$e^{\Phi_o}={f_1^2\over b_o^2R_o^2\epsilon^2(R_o)}c_oe^{-\mu(R_o)},\quad 
\Phi(R_o)=\Phi_o\eqn\qq$$
$$a(r)=(c_oe^{2\mu(R_o)}-1+f_o)e^{-\Phi_o}\eqn\qq$$
where $\epsilon (R_o)=\rho_o+p_{ro}$ and $b(t_i)=b_o$, $c_o,f_o,f_1$ 
are constants. Free constant $c_o>0$ is such that $a(r)>0$ for all allowed 
$r$. The functional form of $\Phi(R)$ is given by
$$e^{\Phi(R)}={f_1^2\over b_o^2R^2\epsilon^2(R)}c_oe^{-2\mu(R)},\quad 
\eqn\qq$$
Note that $\Phi(R)$ is a $C^2$ function of $R$ for all $R> 0$
and $a(r)$ is also a
$C^2$ function of $r$ within the cloud. These conditions further imply
that $e^{\Theta}$ is also a $C^2$ function of $R$ and $t$ for $R> 0$.
Differentiating (53) with respect to $r$ we get
$$R'= {a'(r)e^{\Phi}\over (e^{\Theta}),_R-(e^{\Phi}),_R}\eqn\qq$$
At $t=t_i$ $R'_o> 0$, and so from the above equation it follows 
that $R'$ could
vanish only when either $e^{\Phi}=0$ or at points where $e^{\Theta}
\rightarrow\infty$. Since both $e^{\Phi}>0$ and $e^{\Theta}>0$ are $C^2$ 
functions for $R>0$, it follows that $R'$ could vanish only at $R=0$.
Thus for $t< t_s$, we have $R'>0$.

The inner shell of the matter is labeled by $r=0$ and the outermost 
shell by $r=r_b$.
At $t=t_i$ the density and pressures are finite throughout the cloud and the
Kretchmann scalar is bounded at $t=t_i$. The singularity appears first 
when the innermost shell collapses to zero volume in the sense that
$R(t_s,0)=0$. The time of collapse of further successive shells to
singularity is given by $R(t_s(r),r)=0$. 
The weak energy condition is satisfied if $f(t)>0$ and $\dot f(t) >0$.
The interesting feature of such an example is that at the singularity 
$R=0$ the mass function $F=0$, and as such for $f(t)<1$ we have $R>F$ and the
entire singularity curve is not covered by the apparent horizon. Thus we have 
a naked singularity where not just the central point but the entire 
singularity curve is visible. On the
other hand, for $f(t)>1$ the singularity is covered and we have a black hole.
Of course, in either case, the mass function $F$ vanishes on the entire
singularity curve and so this is a ``massless'' singularity in that sense, 
but what is important is this example illustrates the possibilities 
inherent within the Einstein field equations towards the determination of
the final fate of gravitational collapse of a massive matter cloud.

\bigskip

\noindent{\bf 6. CAUCHY PROBLEM AND THE INITIAL DATA}

In previous sections different classes of evolutions for the initial data
leading to either a black hole or a naked singularity were discussed.
We shall make now some general comments on all possible dynamical evolutions
of a given regular initial data in the context of the occurrence of the
singularities, naked or otherwise, for general matter fields. This 
is related in a way to the Cauchy problem in the context of general relativity.
The difference, however, is that while most discussions on Cauchy problem
in relativity avoid singularities, our purpose is precisely to examine the 
possible evolutions of an initial data into a spacetime singularity,
in terms of its visibility or otherwise.

We pointed out in [10] that a naked singularity would occur generically in
spherically symmetric collapse situations. The point of emphasis here would be
to examine this issue in the context of a given arbitrary initial data. 
We examine if the in built freedom of choice in selection of an equation of 
state in general relativity is sufficient enough to allow generic evolutions 
of an arbitrary initial data into a naked singularity or a black hole subject 
to the energy conditions being satisfied throughout the evolution. The basic 
issue is, firstly, given an arbitrary physically reasonable initial data 
describing the state of the matter at the onset of gravitational collapse,
does general relativity generically allow the evolution of this data into
a singularity, subject to the energy conditions being satisfied throughout,
and if so, do all these evolutions have to necessarily end either as black 
hole or a naked singularity, or is there still some degree of freedom 
available in the field equations which would allow the formation of either 
as desired.

To address this issue along the lines of [10], the field equations given 
in equations (7) to (11) can be put in the following form if one uses $R$ and 
$t$ as variables instead of $t$ and $r$,
$$ \rho={F,_R\over k_0R^2}\quad p_r=-{F,_R\over
k_0R^2}-{F,_t\over R^2T}\eqn\qq$$
$$e^{2\nu}=\left({Q,_R\over F,_{RR}+2p_{\theta}}\right)^2\eqn\qq$$
$$\dot R=-{Q,_R\over F,_tQ(F,_{RR}+2p_{\theta}R)}\eqn\qq$$
$$\left({Q^2(1-{F\over R})+F,_t\over F,_t(F,_{RR}+2
p_{\theta}R)}\right)Q,_{RR}-Q,_t=f_1(R,t,Q,Q,_R)
\eqn\qq$$
$$ H=G-1+{F\over R}\eqn\qq$$
where $,_R$ and $,_t$ represent partial derivatives with respect to
$R$ and $t$ keeping $t$ and $R$ as constants respectively, and we have put
$$Q={F,_t\over \sqrt{G-1+{F\over R}}},\quad p=F,_{RR}+2p_{\theta}R\eqn\qq$$
$$f_1(R.t,Q,Q,_R)={Q,_R\over Qp}\left(Q,_R+Q^2({G[p^2
\dot F],_R\over 2p^2\dot F^2}+{[{F\over R}],_R\over {F\over R}})
+{\dot F,_R\over 2\dot F}
\right ) +{Q^3\over 2R}-{\ddot FQ\over 2\dot F}
\eqn\qq$$

The mass function $F$ and the tangential pressure $p_{\theta}$ 
are free 
functions of $R$ and $t$, and give the distribution of matter throughout 
the spacetime. As usual $R$ at the initial $t=0$ is scaled as $R(0,r)=r$
using the scaling freedom available. Knowing
$F(R,t)$ and $p_{\theta}(R,t)$ one integrates the second order parabolic 
partial differential equation (65) with appropriate initial and boundary
data to obtain $Q(R,t)$. Then $e^{2\nu}$ is immediate from equation (63).
Equation (64) is then integrated for $R(t,r)$ with the initial condition 
$R(0,r)=r$. Equations (66) is then solved for the remaining unknown $\psi$.
The initial data, as pointed out in section 2, consists of four
independent functions of coordinate $r$ at an initial spacelike slice $t=0$ 
when the collapse commences. These
are basically the density distribution $\rho_o(r)$, radial pressure 
$p_{ro}(r)$, and tangential pressure $p_{\theta o}(r)$, which describe the 
state of matter at the onset of collapse, and the velocity distribution 
function $V_o=V_o(r)=-\dot R(0,r)$, which describes the radial velocities 
of the spherical shells towards the center. Thus, for an arbitrary initial 
data at $t=0$ we have
$$F(R,0)=F(r,0)=F_o(r)=\int_0^r{r^2\rho_odr},\quad F,_t(R,0)=F_1(R)=F_1(r)=
r^2p_{ro}V_o\eqn\qq$$
$$p_{\theta}(R,0)=p_{\theta}(r,0)=p_{\theta 0}(r)\eqn\qq$$
The functions $F$ and $p_{\theta}$ are free functions to be chosen, 
subject to their initial values at $t=0$. The functional behavior
of $F$ and $p_{\theta}$ for $t>0$ is completely free except that they should 
at least be $C^2$, and such that energy conditions are satisfied throughout 
the collapse, i.e. $\rho\ge0, \rho+p_i\ge0$. This will be the case when
$$ F,_R\ge 0,\quad  F,_t\ge 0,\quad p_{\theta}\ge -{F,_R\over k_oR^2}
\eqn\qq$$
Thus energy conditions are satisfied for all the evolutions where
the mass function is a monotone increasing function of both $R$ and $t$, with
non-negative tangential pressures.

We therefore consider only those sets of arbitrary functions 
$F(R,t)$ and $p_{\theta}(R,t)$ for which $F\ge 0, F,_R>0, F,_t>0$,
$p_{\theta}>-\rho$, and $p=F,_{RR}+2p_{\theta}R>0$ everywhere except perhaps 
at the singularity $R=0$ where it could vanish. Note that in case
$p=0$ the parabolic partial differential equation (65) is replaced
by an second order ordinary differential equation for $T=\dot R$ with
$Q=c(t)F,_t$, and all our conclusions here would again apply in this
case. We would further consider only those sets of functions for which 
$f_1$ in equation (68) is at least $C^0$ function of its arguments. From 
equation (14) which gives the expression for the Kretchmann scalar, it is 
clear that at the initial time $t=0$ it is finite since the initial data in 
the form of density, pressures, and velocity functions are all at least
$C^2$ functions, and therefore at $t=0$ we have $F(R,0)=R^3h(R)$, where $h(R)$
is at least $C^1$ function of $R$. Therefore, for all sets $F(R,t)$
a shell-focusing singularity $R(t_s,r)=0$ would develop at a later time 
$t=t_s(r)$ if $F\simeq R^n, n<3$.
Similar situations would occur when we have 
$R'(t,r)=0$, and a shell-crossing singularity would occur.
Thus for a given set of arbitrary initial data of matter and
velocity functions at the initial spacelike hypersurface, there
are infinite many evolutions satisfying the energy conditions, characterized 
by the free functions $F$ and $p_{\theta}$, resulting into a singularity.
The question that needs to be answered is for a particular evolution
determined by a particular choice of functions $F$ and $p_{\theta}$,
would there be a naked singularity or a black hole.

Note that the second order partial differential equation above for $Q$ is
a quasi-linear parabolic equation in variables $R$ and $t$. The coefficient
of $Q,_{RR}$ is positive and at least $C^0$ (depending upon the choice
of $F$ and $p_{\theta}$), while that of $Q,_t$ is $-1$.
Furthermore, the driving term $f_1(Q,R,t,Q,_R)$ is also at least $C^0$ 
function of its argument. The initial value of $Q(R,0)=Q_o(R)$ at $t=0$ is 
fixed by the initial data as mentioned earlier. Thus we have a well posed
Cauchy problem with the boundary condition at some $t=t_s$ still free to 
choose. The quasi-linear parabolic partial differential equations, such
as (74), have been studied quite extensively for existence of solutions
for an initial condition at $t=0, Q(R,0)=Q_o(R)$, and the boundary condition
at some $t=t_s$, and the solutions do exist with quite general form of 
boundary conditions $Q(R,t)=g(R,t)$. Therefore, for all sets of functions
$(F,p_{\theta})$ for which the coefficient of $Q,_{RR}$ in equation (65)
is positive, and together with
$f_1$ is at least $C^0$ function of its argument,  
the solution with quite general boundary conditions would exist which
would not only evolve from a given initial data but also satisfy the energy 
conditions throughout evolution. This is the case as shown in examples
considered in earlier sections. The point is that with the choice of a 
particular initial data set, namely $(\rho_o,p_{ro},p_{\theta o}, V_o)$, 
and a particular set of evolutions given by $(F(R,t),p_{\theta}(R,t)$ we
still have the choice of the boundary condition at $t=t_s$ at our disposal. 
The freedom left in the form of the choice of $Q(R,t_s)$, once chosen,
completes the solution of the field equation and it is this choice which 
effects the nature of the singularity occurring at $t=t_s, R=0$ in terms of
being naked or otherwise. This is precisely what happens in the classes
considered in earlier sections
where an evolution from an arbitrary initial matter data can form either a
black hole or a naked singularity.

To clarify this, let us briefly consider the radial null geodesics 
equation, which is,
$${dt\over dr}=e^{\psi-\nu}\Rightarrow {dR\over dt}={QQ,_R(1-{F\over R})
\over p(\sqrt{\dot F}+\sqrt{\dot F+Q^2(1-{F\over R})}}\equiv U(X,t)\eqn\qq$$
where $X=R/t-t_s$. Since at the
initial $t=0$, $R=r$ and $R'=1$, and $\dot R<0$ throughout the spacetime,
the singularity would occur at a later time $t=t_s>0$. From equation
(74) it follows that for a choice of functions $(F,p_{\theta})$ such that
$\dot F>0, p>0$ throughout the spacetime, $Q$ has at least two continuous
$R$-derivatives and that $p$ has at least one continuous R-derivative, 
then $R'>0$ and is finite throughout the spacetime if it starts that way,
except possibly at the singularity $R=0$. Thus in such cases shell-crossings 
would not occur, and the first shell-focusing singularity would occur at the
center at $r=0, t=t_s$. Note that in case $F(0,t_s)>0$ the singularity
would be covered. The singularity would be naked and there would be 
outgoing radial null geodesics if the equation $V(X)=U(X,t_s)-X=0$, where
$$X_o=\lim_{(R\to 0,t\to t_s)}\left({R\over t-t_s}\right)=
\lim_{(R\to 0,t\to t_s)}\left({dR\over dt}\right)=U(X_o,t_s)\eqn\qq$$
has a real positive root $X=X_o$. In the case when there are no real positive 
roots for $V(X)=0$, then the singularity is covered. The value of the function
$Q(R,t_s)$ which appears in the root equation above thus becomes important.
For all functions $Q(R,t_s)$ chosen such that the equation has a root gives
rise to the naked singularity, the choice otherwise leads to a black hole.
$Q(R,t_s)$ is precisely the boundary condition in equation (65) which one has
a freedom to choose for obtaining a particular solution. Therefore, as 
illustrated by the two examples in the preceding sections, there
would be a non-zero measure set of evolutions from an arbitrary
initial data specified at the onset of the collapse, which could
always develop into a singularity, naked or covered.

\bigskip

\noindent{\bf 7. CONCLUDING REMARKS}

We considered here various classes of collapsing evolutions from 
arbitrary initial data and it is shown that these can be chosen so as to 
form either a black hole or a naked singularity. Such an initial data, 
prescribed in terms of the initial density and pressure profiles of the cloud,
and the initial velocities of the matter shells, can evolve into either of 
these outcomes depending on the evolution chosen. In other words, there are 
permissible classes of evolutions, subject to the energy conditions and 
other conditions ensuring the physical reasonability, which give rise to 
either a black hole as the final product of collapse, or a naked singularity.

It appears that the scenario presented above is repeated in nearly all 
known exact physically reasonable solutions of the field equations 
describing spherically symmetric gravitational collapse.  
In the known classes, such as radiation collapse [5],
dust [11], perfect fluids and more general forms of matter [7], and
massless scalar field collapse [2], singularities both naked or covered do 
occur. Therefore, from the point of view of cosmic censorship, the 
important question currently is not 
the occurrence of naked singularities but whether these occurrences are 
permitted by the general relativity on a generic enough basis, and whether 
they are stable according to a suitably defined stability criteria.
The issue of genericity is significant from the point of view that 
these phenomena could be impossibly
rare in the space of solutions of the general theory of relativity. The known
literature does not indicate this clearly, because for all known 
important classes of solutions in the spherically symmetric gravitational 
collapse both the naked singularities and black holes occur for wide 
ranging situations.

An important indicator in this direction, as pointed out earlier, 
is the imploding Vaidya model, where the singularity is naked 
for $\lambda\le 1/8$ and a black hole for $\lambda >1/8$, where the
parameter $\lambda$ represents the initial data in the form of
the rate of mass loss. Thus, the singularities, both naked and covered,
are stable against the pertubation of the parameter $\lambda$, and the point
$\lambda=1/8$ is the critical point indicating the transition from one
phase to the other. A similar
situation also occurs in the case of dust collapse where
the pertubation in the initial density or velocity distribution within
a certain domain does not alter the nature of the singularity. The analysis
here in general has a significance in that the nature of the 
singularity is seen to be stable in a certain sense against the pertubation 
of the initial data.

In this sense, we have characterized here wide new families 
of black hole solutions forming in spherically symmetric gravitational 
collapse, without using the assumption of the cosmic censorship hypothesis, 
but in terms of the possible evolutions of the initial data for the 
collapsing object. What we still do not know is the actual measure of each 
of these classes in the space of all possible evolutions which are allowed 
from a given general and arbitrary but physically reasonable initial data set.
This is a problem related closely to the issue of stability of naked
singularities. As is well-known, the stability in general relativity
is a complicated issue because there is no well-defined formulation or
specific criteria to test for stability. Fast evolving numerical codes 
for core collapse models may possibly provide further insights into this
aspect. All the same, these classes appear to be generically arising in the 
collapse models considered here, at least within spherical symmetry,
in that they are not an isolated phenomena but belong to a general
family. Because, given any density and pressure profiles for the cloud, there 
exists an evolution which will lead to either a black hole or a naked 
singularity as desired as the end product 
of collapse. In this sense, both black holes and naked singularities do seem 
to arise generically as the end product for spherically symmetric 
gravitational collapse. Given the complexity of the field equations,
if a phenomena occurs so widely in spherical symmetry, it is not unlikely 
that the same would be repeated in more general situations.

It thus appears from the above considerations that the 
occurrence of singularities, naked or otherwise, is inherent in the theory of
general relativity, and a distinction between these cases may not be possible 
through general relativity alone. What is essential is to examine how the 
perturbations and departures from sphericity would alter 
these conclusions. Some efforts have been made to examine this issue [12]. 
It is also possible that the quantum effects near the naked singularities
may help to preserve some kind of a quantum cosmic censorship, or these
quntum effects could give rise to interesting signatures for naked
singularities [13].

\vfill\eject

\centerline{\bf References}

\item{[1]} I. H. Dwivedi and P. S. Joshi, Class. Quant. Grav. 14, p.1223
(1997); P. S. Joshi and T. P. Singh, Phys. Rev. {\bf 51} (1995) 6778;
T.P. Singh and P. S. Joshi, Class. Quantum Grav., 13, p.559 (1996).

\item{[2]} R. M. Wald, {\it Gravitational Collapse and Cosmic Censorship},
gr-qc/9710068; to appear in {\it The Black Hole Trail}, (ed. B. Iyer).

\item{[3]} see e.g. R. Penrose, in {\it Gravitational Radiation  and 
Gravitational Collapse}, Proceedings of the IAU Symposium, edited by C.
DeWitt-Morette, IAU Symposium No 64 (Reidel, Dordtecht, 1974); 
R. Hagedorn, Nuovo Cimento A56, 1027 (1968).

\item{[4]} S. W. Hawking and G. F. R. Ellis, {\it The large scale
structure of space-time}, Cambridge University Press, Cambridge (1973).

\item{[5]} P. S. Joshi, {\it `Global aspects in gravitation and cosmology'},
Clarendon Press, OUP, Oxford (1993).

\item{[6]} R. Penrose, Riv. Nuovo Cimento {\bf 1} (1969) 252; in 
{\it General Relativity, An Einstein Centenary Survey}, ed. S. W. Hawking 
and W. Israel, Cambridge Univ. Press, Cambridge, London (1979) 581.

\item{[7]} A. Ori and T. Piran, Phys. Rev. D 42, p.1068 (1990); 
P. S. Joshi and I. H. Dwivedi, Commun. Math. Phys. 146, p.333 (1992); Lett. 
Math. Phys. 27, p.235 (1993); K. Lake, Phys. Rev. Lett. 68, p.3129 (1992); 
P. Szekeres and V. Iyer, Phys. Rev. D 47, p.4362 (1993); F. Cooperstock, 
S. Jhingan, P.S. Joshi, T. P. Singh, Class. Quant. Grav. 14, p.2195 (1997); 
G. Magli, Class. Quant. Grav. 14, p.1937 (1997); T. P. Singh and L. Witten, 
Class. Quant. Grav. 14, p.3489 (1997); G. Magli, gr-qc/9711082.

\item{[8]} C. J. S. Clarke, {\it `Analysis of space-time singularities'}, 
Cambridge University Press, Cambridge (1993).

\item{[9]} S. Jhingan, P. S. Joshi and T. P. Singh, Class. Quant. Grav. 
13, p.3057 (1996).

\item{[10]} I. H. Dwivedi and P. S. Joshi, Commun. Math. Phys. 166, 
p.117 (1994).

\item{[11]} D. M. Eardley and L. Smarr, Phys. Rev. D {\bf 19}, (1979) 2239; 
D. Christodoulou, Commun. Math. Phys. {\bf 93} (1984) 171; R. P. A. C. 
Newman, Class. Quantum Grav. {\bf 3} (1986) 527; B. Waugh and K. Lake, Phys. 
Rev. D {\bf 38} (1988) 1315; I. H. Dwivedi and P. S. Joshi, Class. Quant.
Grav., 9, L69 (1992); P. S. Joshi and I.H. Dwivedi, Phys. Rev. {\bf D47} 
(1993) 5357.

\item{[12]} P. S. Joshi and A. Krolak, Class. Quant. Grav. 13, p. 3069 (1996); 
H. Iguchi, K. Nakao, T. Harada, Phys. Rev. D, 15 April, 1998; A. Chamorro,
R. Gregory, and J. Stewart, Proc. Roy. Soc. Lond. A413, 251 (1987).

\item{[13]} L. Ford and L. Parker, Phys. Rev. {\bf D17} (1978) 1485;
W. A. Hiscock, L. G. Williams and D. M. Eardley, Phys. Rev. {\bf D26} (1982) 
751; Cenalo Vaz and Louis Witten, Phys. Letts. {\bf B325} (1994) 27; 
Class. Quant. Grav. {\bf 12} (1995) 1; {\it ibid.} {\bf 13} (1996) L59; 
Nucl. Phys. {\bf B487} (1997) 409; S. Barve, T. P. Singh, C. Vaz, L. Witten,
gr-qc/9802035; C. Vaz and L. Witten, gr-qc/9804001.

\end